\documentclass[%
 reprint,
 superscriptaddress,
 preprintnumbers,
 amsmath,amssymb,
 aps,
]{revtex4-1}

\usepackage{graphicx}% Include figure files
 \usepackage{dcolumn}% Align table columns on decimal point
\usepackage{bm}% bold math
\usepackage{color}
\usepackage[normalem]{ulem}%Added by Martin, to cross out words with the command \sout{}

\usepackage{hyperref}% add hypertext capabilities
\newcommand{\fref}[1]{Fig.~\ref{fig:#1}} 
\newcommand{\eref}[1]{Eq.~\eqref{eq:#1}}

\newcommand{\cref}[1]{Chapter~\ref{ch:#1}}
\newcommand{\tref}[1]{Table~\ref{tab:#1}}

\newcommand{\nnl}{\nonumber \\}

\newcommand{\beq}{\begin{equation}} 
\newcommand{\eeq}{\end{equation}} 
\newcommand{\ba}{\begin{array}}  
\newcommand{\ea}{\end{array}} 
\newcommand{\bea}{\begin{eqnarray}}  
\newcommand{\eea}{\end{eqnarray} }  
\newcommand{\be}{\begin{eqnarray}}  
\newcommand{\ee}{\end{eqnarray} }  
\newcommand{\bal}{\begin{align}}
\newcommand{\eal}{\end{align}}   
\newcommand{\bi}{\begin{itemize}}  
\newcommand{\ei}{\end{itemize}}  
\newcommand{\ben}{\begin{enumerate}}  
\newcommand{\een}{\end{enumerate}}  
\newcommand{\bc}{\begin{center}}
\newcommand{\ec}{\end{center}} 
\newcommand{\bt}{\begin{table}}
\newcommand{\et}{\end{table}}  
\newcommand{\btb}{\begin{tabular}}
\newcommand{\etb}{\end{tabular}}

\newcommand{\cO}{{\mathcal O}}

\newcommand{\mev}{\mathrm{MeV}}
\newcommand{\gev}{\mathrm{GeV}}

\newcommand{\eps}{\epsilon}

\begin{document}

\title{21cm absorption signal from charge sequestration}

\author{Adam Falkowski}
  \email{adam.falkowski@th.u-psud.fr}
\affiliation{Laboratoire de Physique Th\'{e}orique, CNRS, Univ. Paris-Sud, Universit\'{e}  Paris-Saclay,  91405 Orsay, France}
\author{Kalliopi Petraki}%
  \email{kpetraki@lpthe.jussieu.fr}
\affiliation{Laboratoire de Physique Th\'{e}orique et Hautes Energies (LPTHE), UMR 7589 CNRS \& Sorbonne Universit\'{e}, 4 Place Jussieu, F-75252, Paris, France}
\affiliation{Nikhef, Science Park 105, 1098 XG Amsterdam, The Netherlands}

\newcommand{\adam}[1]{{\color{black} #1}}
\newcommand{\kallia}[1]{{\color{red} #1}}


\begin{abstract}

The unexpectedly strong 21cm absorption signal detected by the EDGES experiment suggests  that the baryonic gas was colder at redshift $z\sim 17$ than predicted in the standard scenario. We discuss a mechanism to lower the baryon temperature after recombination. We introduce a stable, negatively-charged particle with a non-negligible cosmological abundance, such that the universe remains charge-neutral but the electron and proton numbers are no longer equal. The deficit of electrons during recombination results in an earlier decoupling of the baryon gas temperature from that of the cosmic microwave background (CMB). This implies a smaller ratio of the gas and CMB temperature at $z\sim 17$. The parameter space of the mechanism where the 21~cm absorption signal is significantly enhanced is probed by the CMB spectrum, cooling of stars and supernovae, and colliders. Nevertheless, we find  viable regions corresponding to sub-eV or MeV-scale milli-charged particles, or to TeV-scale multi-charged particles.

\end{abstract}

\maketitle

{\bf Introduction.}   
The EDGES observation of the 21cm absorption signal \cite{Bowman:2018yin} opens a new window on the early universe, during the so-called dark ages and cosmic dawn epochs,  allowing us to test the standard $\Lambda$CDM scenario and place novel constraints on hypothetical particles and interactions beyond the Standard Model (SM). 
The amplitude of the signal is described by the formula {\cite{Madau:1996cs,Zaldarriaga:2003du}}  
\beq
\label{eq:T21}
T_{21}[{\rm K}] \approx  {0.035} \left(1  -  {T_\gamma(z)\over T_s(z)}  \right )  \sqrt{1 + z \over 18} , 
\eeq 
where $T_\gamma$ is the temperature of the CMB radiation, $T_s$ is the spin temperature of the hydrogen gas describing the relative occupation number of the singlet and triplet states,  and we used  $\Omega_b =  0.0449$, $\Omega_m = 0.3156$ \cite{Ade:2015xua}.  
One expects that, during  the relevant epoch,  the spin temperature is coupled to the kinetic gas temperature: $T_\gamma \gg T_s \gtrsim T_g$ for $z \in [15,20]$. 
The standard $\Lambda$CDM scenario predicts  $T_\gamma(17) \approx  49$~K, $T_g(17) \approx  6.8$~K \cite{AliHaimoud:2010dx},  which implies $T_{21} \gtrsim -0.2$~K. 
On the other hand, EDGES finds $T_{21} \approx -0.5^{+0.2}_{-0.5}$~K, where the quoted uncertainty is $99\%$~CL,  which corresponds to a $3.8~\sigma$ deviation from the  $\Lambda$CDM  prediction.

From \eref{T21} it is apparent that the absorption signal observed can be enhanced either by modifying the CMB spectrum~\cite{Fraser:2018acy,Pospelov:2018kdh} or by decreasing the baryon gas temperature at $z \sim 17$. 
The latter can be achieved in the presence of interactions between baryons and dark matter (DM) \cite{Dvorkin:2013cea,Tashiro:2014tsa}, however concrete realization of that idea run into severe observational constraints \cite{Barkana:2018lgd,Munoz:2018pzp,Berlin:2018sjs,Barkana:2018qrx}. 
In this letter we pursue a different path to cooling the baryon gas during the dark ages. 
In the standard scenario, the baryon temperature decouples from that of the CMB around $z\sim 200$ because the Compton scattering rate of CMB photons on free electrons falls  below the critical value. 
After the decoupling, the baryon gas cools faster with the decreasing redshift,  as $T_g \propto (1+ z)^2$, compared to $T_\gamma \propto (1+z)$. 
The Compton scattering  rate is proportional to the electron ionization fraction $x_e$. 
Therefore, when $x_e$ is {\em decreased} compared to the standard evolution, the decoupling occurs {\em earlier}, and baryons are {\em colder}  at the cosmic dawn. 

In the standard scenario the number densities $n_e$, $n_p$ of free electrons and protons are equal at all times, ensuring the charge neutrality of the universe. We propose that there exists another stable particle with a negative electric charge and with a non-negligible abundance {(at least) around the time of recombination.}   
Denoting our particle as $X$, and its electric charge $-\eps_X$, the  charge neutrality  condition becomes:
\beq
\label{eq:master}
x_p = x_e + \eps_X r_X,  
\eeq 
where $x_p \equiv n_p/n_B$, $x_e \equiv n_e/n_B$, $r_X \equiv n_X/n_B$ (more generally, $r_X = (n_X - n_{\bar X})/n_B$), and $n_B$ is the baryon number density. The presence of the new charged component affects the {ionization} history. We will show that $\eps_X r_X \gtrsim 10^{-4}$ leads to   a significant suppression of $x_e$ after recombination. This, in turn, implies an earlier decoupling of the gas from the CMB photons, a smaller value of $T_g$ at the cosmic dawn, and thus a stronger 21~cm absorption signal. 

{\bf Ionization fraction and gas temperature.} 
We study the ionization history in a universe where $n_e \neq n_p$ using the 3-level atom model~\cite{Peebles:1968ja,Zeldovich:1969en,Senatore:2008vi}. 
In this approximation, recombination of hydrogen proceeds through the 2S and 2P levels, while higher excited levels are ignored.  
Furthermore, one assumes most of the baryons are in the form of protons and ground-state hydrogen atoms,  $n_B \approx n_p + n_H \approx n_p + n_{1S}$, while helium and heavier elements are ignored.  
Finally, we assume the CMB is not affected by new physics, thus  $T_\gamma (z) = T_0 (1 + z)$, $T_0 = 2.725$~K.  
Adapted to our case, the evolution equation for the ionization fraction  takes the form 
\bea 
\label{eq:RECO_dxe}
{\partial x_e \over \partial z}  & = & 
{ \left [ \alpha_B(T_g) n_B x_e x_p - \beta_B(T_\gamma)  e^{-E_{12}/T_\gamma} (1 -x_p) \right ]
 \over 
 (1 + z) H }
 \nnl & \times & 
 {\left [ 1 + \left (1 - x_p \right )  \Lambda_{2S \to 1S}  {\pi^2  \over H E_{12}^3}  \right ] 
\over  
 \left [ 1
  + \left (1 - x_p \right ) \left ( \Lambda_{2S \to 1S}  +\beta_B(T_\gamma)  \right )  {\pi^2  \over H E_{12}^3}   \right ]  },  
\eea 
with 
\begin{align*}
\alpha_B(x) 
&\approx {9765.88 \over \gev^2} {4.309 (x/10^4{\rm K})^{-0.6166} \over 1 + 0.6703 (x/10^4 {\rm K})^{0.53}} ,
\\
\beta_B(T) 
&= \alpha_B(T) e^{-E_2/T} \left ( m_e T \over 2 \pi \right )^{3/2} ,
\end{align*}  
$E_2 = 3.4$~eV, $E_{12} =  10.2$~eV,  $\Lambda_{2S \to 1S} \approx 5.4 \times 10^{-15}$~eV. 
We use the electron mass  $m_e \approx 511$~keV,  and the Hubble function $H  = \sqrt{\Omega_m} H_0 (1 + z)^{3/2}$, $H_0  \approx 1.43\times 10^{-33}$~eV.
The proton and electron ionization fractions are related as in \eref{master}. 

We solve \eref{RECO_dxe} together with the one describing evolution of $T_g$ \cite{AliHaimoud:2010ab}: 
\begin{subequations}
\label{eq:RECO_dtg}
\begin{align}
{\partial T_g \over \partial z}  
&= {2 T_g - \gamma_C (T_\gamma - T_g) \over 1+ z } , 
\\
\gamma_C   
&=   {8 \sigma_T a_r T_\gamma(z)^4 \over  3 H m_e} {x_e \over 1 + f_{\rm He} + x_e} , 
\end{align}
\end{subequations}
where {the helium fraction is} $f_{\rm He} \approx 0.08$, $\sigma_T =  {8 \pi \over 3}{\alpha^2 \over m_e^2}$ is the Thomson cross section, and $a_r= \pi^2/15$ is the radiation constant. 
Besides electrons, the $X$ ions may also mediate photon-gas interactions that can heat up the gas. The ratio of the energy transfer by  $X$  over that by the electrons is 
$n_X \sigma_T^{X\gamma} / n_e \sigma_T^{e\gamma}  
\propto \epsilon_X^4 (r_X/x_e) (m_e^2/m_X^2)$, which turns out to be much less than 1 in the entire parameter space allowed by other constraints (cf.~next section). We therefore neglect the energy transfer by $X$ in our analysis.  
Moreover, we assume here that no other effect, such as baryon-DM scattering, significantly affects the baryon temperature.

We solve the system of \eref{RECO_dxe} and  \eref{RECO_dtg} numerically for different values of  $\eps_X r_X$. 
The results are shown in \tref{RECO_qrx2} and in \fref{reco}. 
For the vanishing charge asymmetry, our approximations reproduce within $10\%$ accuracy the baryon temperature obtained in  fully-fledged simulations \cite{AliHaimoud:2010dx}, which is completely satisfactory for our purpose.  
Once $\eps_X r_X \gtrsim 10^{-4}$, $x_e$ becomes suppressed compared to the standard value (while $x_p$ is slightly enhanced),  which accelerates the decoupling of the hydrogen gas and thus lowers the gas temperature at the cosmic dawn.  

\begin{table}
\bc 
\begin{tabular}{|c|c|c|c|}
\hline
$\eps_X r_X$ &  $x_e^{\infty}(17)$ &  $x_p^{\infty} (17)$  & $T_{g}(17)$ [K]   \\ \hline 
$0$ & $1.5 \times 10^{-4}$ & $1.5 \times 10^{-4}$ & 6.2 \\ \hline  
$10^{-5}$ & $1.4 \times 10^{-4}$& $1.5 \times 10^{-4}$ & 6.0  \\ \hline
$10^{-4}$ & $9.9 \times 10^{-5}$ & $2.0 \times 10^{-4}$ &5.5  \\ \hline 
$3 \times 10^{-4}$ & $3.8 \times 10^{-5}$ & $3.4 \times 10^{-4}$ & 4.5  \\ \hline 
$5 \times 10^{-4}$ & $1.1 \times 10^{-5}$ & $5.1 \times 10^{-4}$ & 3.7  \\ \hline 
$7 \times 10^{-4}$ & $2.5 \times 10^{-6}$ & $7.0 \times 10^{-4}$ & 2.5  \\ \hline 
\end{tabular}
\ec
\caption{
\label{tab:RECO_qrx2}
Free electron and proton fractions and the baryon gas temperature at $z = 17$ for different contributions of particle $X$ to the charge budget of the universe. 
}
\end{table}

\begin{figure}[htb]
\bc 
\includegraphics[width=0.85\linewidth]{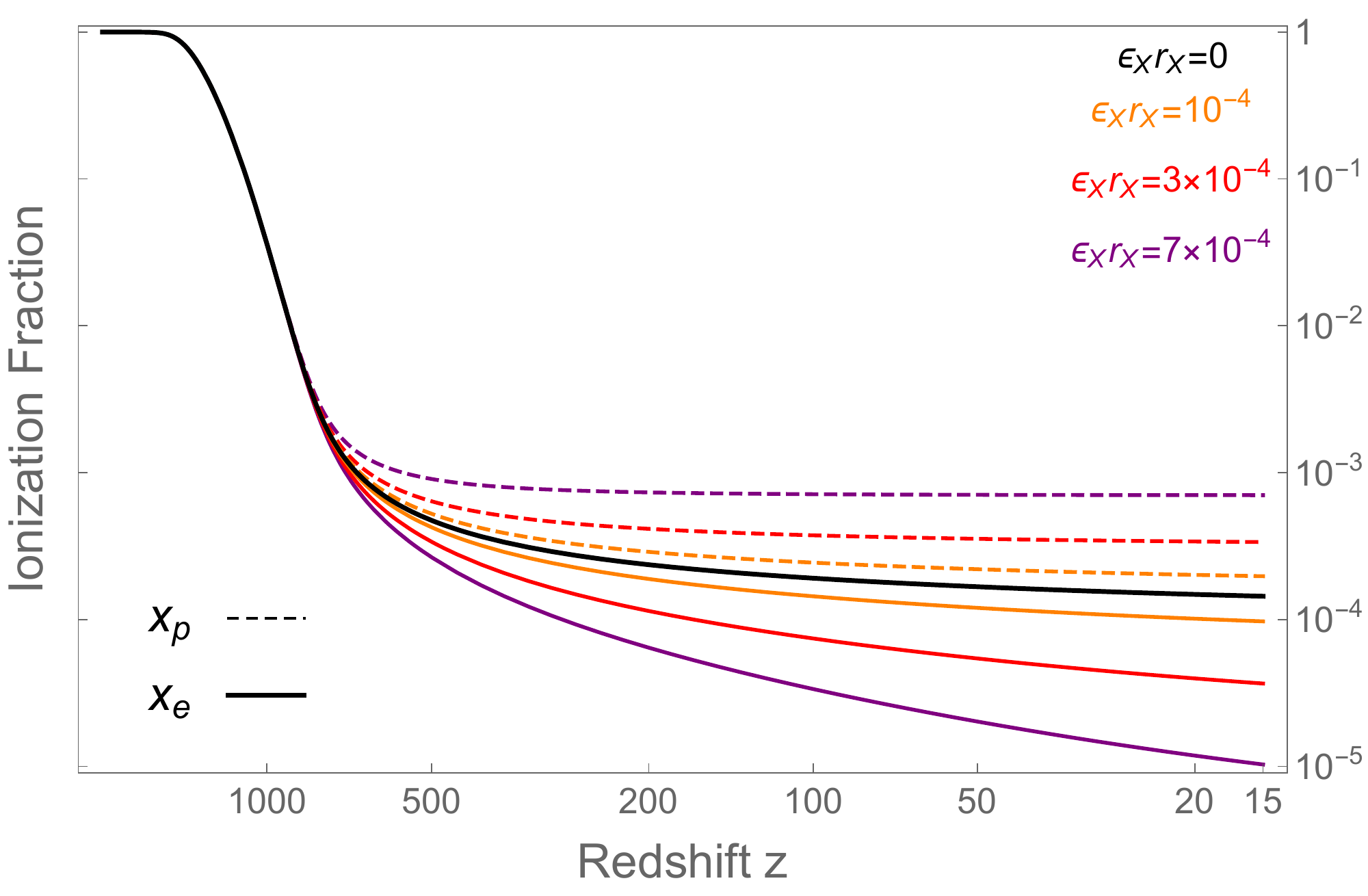}
\ec 
\bc 
\includegraphics[width=0.85\linewidth]{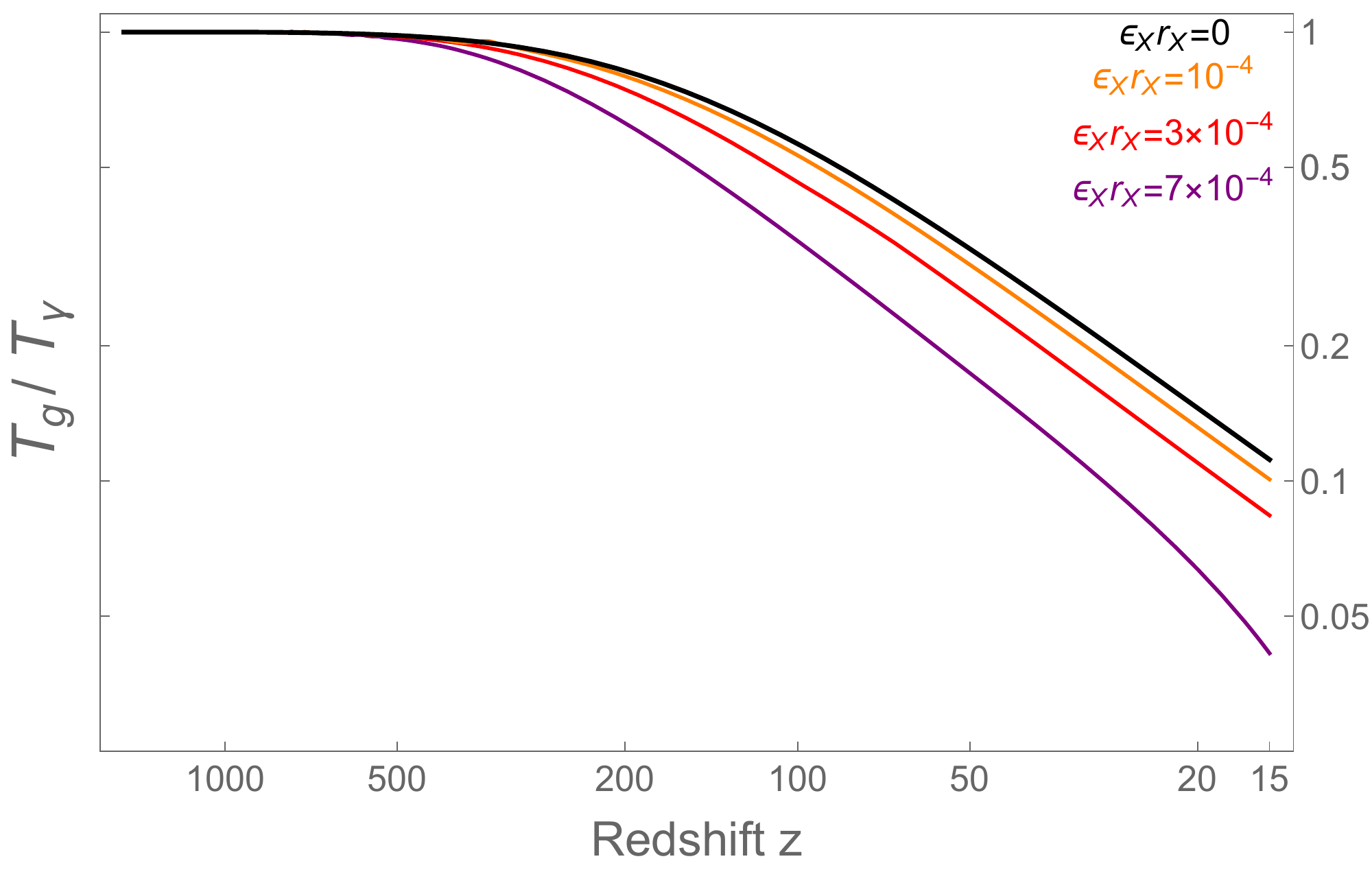}
\ec
\caption{\label{fig:reco}
Evolution of the ionization fraction  (top)  of electrons (solid) and protons (dashed) and of the ratio  hydrogen to CMB  temperatures (bottom) as function of redshift $z$. Lines of different colors correspond to different values of $\eps_X r_X$: $0$ (black), $10^{-4}$ (orange), $3 \times 10^{-4}$ (red),  $7 \times 10^{-4}$ (purple).}
\end{figure}

{\bf Constraints.} 
We now discuss the constraints on a stable particle with mass $m_X$ and electric charge $-\eps_X$ contributing  $\eps_X r_X \sim 10^{-4}$ to the charge budget of the universe. We shall assume that no significant number of $X$ antiparticles is present today (we comment on cosmological scenarios realizing this feature in the next section). Then, the contribution of the $X$ species to the DM density, $f_X \equiv \Omega_X  / \Omega_{\rm DM}$, is  
\beq 
f_X = r_X {m_X \over m_p}  {\Omega_b \over \Omega_{\rm DM}}
\simeq \left ( \eps_X r_X \over 10^{-4}  \right ) 
\left (10^{-6} \over  \eps_X \right ) \left ( m_X \over 50~\mev \right ) . 
\label{eq:fX}
\eeq 
We do not insist that $X$ constitutes all of DM, and only require $f_X \leqslant 1$, which selects a viable  region of the $m_X$-$\eps_X$ parameter space  where the mass-to-charge ratio is small enough. Moreover, the CMB anisotropies impose $\Omega_X \lesssim 0.002$ if the $X$ particles are tightly coupled to the plasma, which occurs for $\eps_X^2 (\mu_{X,p}^{1/2} + \mu_{X,e}^{1/2})/m_X \geq 5\times 10^{-11} \gev^{-1/2}$~\cite{Dolgov:2013una}, where $\mu_{X,i} = m_X m_i/(m_X + m_i)$.  Furthermore, independently of the cosmological history, there are stringent constraints from collider and beam dump experiments, and from cooling of stars and supernovae.
Altogether, these constraints leave only small unexcluded patches in the $\eps_X$-$m_X$ plane, as shown  in \fref{millicharge}.  
The viable possibilities are the following:
\begin{figure}[htb]
\bc 
\includegraphics[width=0.99\linewidth]{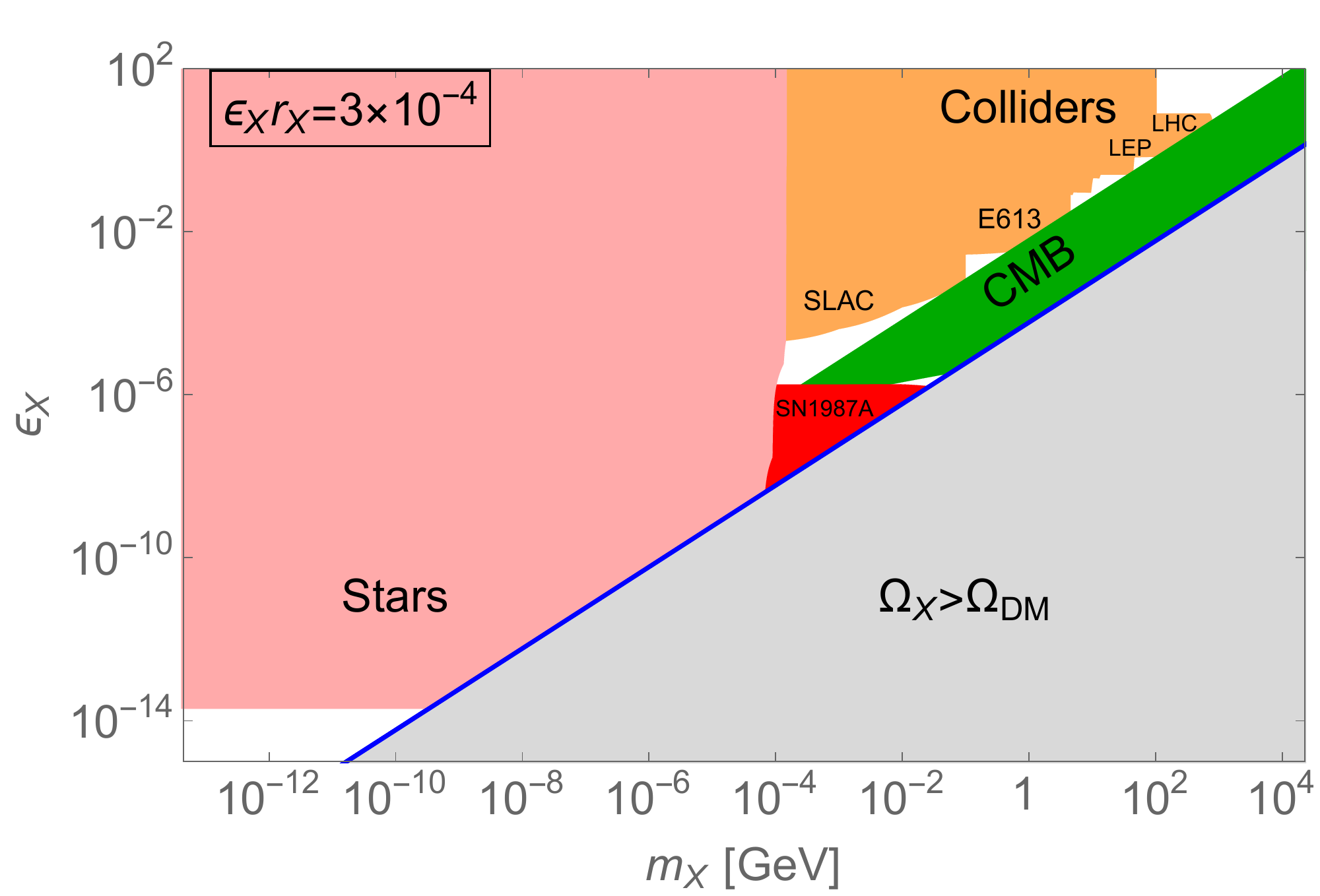}
\ec 
\caption{
\label{fig:millicharge}
Constraints on the parameter space of  charged particles assuming they contribute
$\eps_X r_X = 3 \times 10^{-4}$ to the charge budget of the universe. 
The constraints from millicharge {particle emission in stars} are taken from Fig.~1 of \cite{Vinyoles:2015khy}, see references therein.
For the SN~1987A constraints we use the analysis of Ref.~\cite{Chang:2018rso} assuming the ``fiducial" density profile.
The collider constraints use the results from SLAC \cite{Prinz:1998ua}, E613 \cite{Soper:2014ska}, LEP \cite{Davidson:2000hf}, and LHC \cite{Chatrchyan:2013oca,Aad:2015oga}. 
On the blue line the energy density of non-relativistic particles $X$ accounts for all of DM in the universe. 
}
\end{figure}
\ben 
\item[\#1]
A small region where $m_X \sim (10$-$100)$~MeV, $\eps_X \sim 10^{-6}$, and $X$ makes all or most of the DM. 
\item[\#2]  
A region  with 100~keV$\lesssim m_X \lesssim$~10~MeV, where 
$10^{-4} \lesssim \Omega_X \lesssim 2\times 10^{-3}$ ($f_X < 0.008$) 
and $\eps_X \lesssim  {\rm few} \times 10^{-5}$ to avoid the constraints from the SLAC millicharge experiment \cite{Prinz:1998ua}. For $m_X \in [0.1,100]$~GeV there are other allowed patches with similar properties passing just under the radar of collider experiments. 
\item [\#3]
 {
A particle with large charge, $5 \lesssim \eps_X \lesssim 100$ and large enough mass to avoid the current collider constraints, where again $\Omega_X \lesssim 0.002$ ($f_X < 0.008$). }
\item[\#4] An ultra-light particle, $m_X \lesssim$~eV,  with a tiny charge,  $\eps_X \lesssim 10^{-14}$, that evades the constraints from emission in stars. 
\een 
Case \#1 is the unique where the new particle could be the usual cold DM candidate, however it faces serious challenges. 
In \fref{omegax1} we show a magnification of the relevant parameter space, fixing the parameters to ensure $f_X = 1$ and letting $\eps_X r_X$ vary.
For a given mass, the CMB constraints yield an upper limit on $\eps_X$. 
Most of the remaining parameter space  is excluded by  the recent analysis of cooling of the supernova 1987A via emission of millicharge particles \cite{Chang:2018rso}. 
In that reference, constraints are derived under different assumptions about the density profile, and the one referred to as ``fiducial'' leads to the most conservative upper limit on $\eps_X$.   
Adopting that result leaves some allowed parameter space with $\eps_X r_X$ large enough to significantly affect the baryon temperature during the dark ages. 
On the other hand, other density profiles studied in \cite{Chang:2018rso} leave only the parameter space with  $\eps_X r_X \lesssim 10^{-4}$, which would have a  limited impact on the baryon gas temperature (cf. \tref{RECO_qrx2}).  
Moreover, in this window the photon-mediated elastic $X p$ scattering cross-section is, numerically, 
\beq
\sigma_{Xp} \approx { 1.5 \times 10^{-38}{\rm cm}^2  \over v_{\rm rel}^4} 
\left ( \eps_X \over 2 \times 10^{-6} \right )^2 . 
\eeq  
This cross-section is large enough for the baryon gas to be cooled by scattering off the DM particles \cite{Barkana:2018qrx}, and this effect will dominate over that of the depleted electron number, unless the DM gas is hotter than in the standard scenario.   
All in all, for $m_X \in (10-100)$~MeV, our assumption of the charge asymmetry of $X$ may be an unnecessary complication.

\begin{figure}[htb]
\bc 
\includegraphics[width=0.9\linewidth]{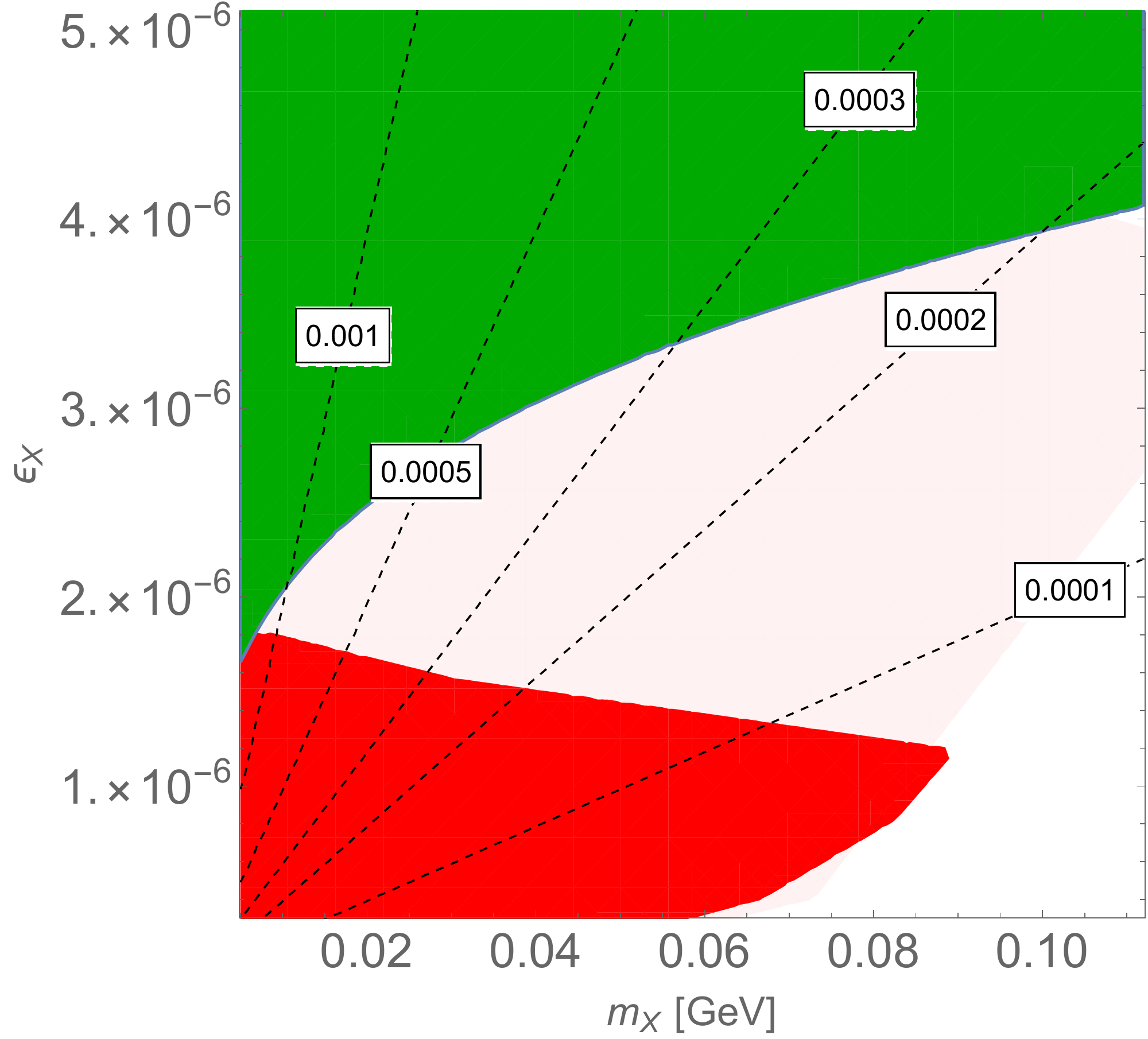}
\ec 
\caption{
\label{fig:omegax1}
Constraints on the parameter space of $\cO(10)$ MeV millicharge particles assuming they constitute all of DM.
We show the CMB constraints (green)~\cite{Dolgov:2013una}, and the more (red) and less (light red) conservative limits  from  cooling of SN~1987A \cite{Chang:2018rso}. 
The labeled contours correspond to different values of $\eps_X r_X$. 
}
\end{figure}

{\bf Discussion.} 
A successful cosmological scenario must
(i)   account for the separation of electric charge between $X$ and the SM species,
(ii)  explain the absence of $X$ antiparticles in today's universe, 
(iii) respect the upper bounds on the radiation density of the universe during BBN and CMB, and 
(iv) ensure that the $X$ particles are sufficiently cold by the time of matter-radiation equality if $f_X \approx 1$.    
Here, we discuss how the above features may be implemented. Concrete particle physics realizations will be presented in a forthcoming publication.

The separation of charge may be achieved, in analogy to asymmetric DM \cite{Petraki:2013wwa,Zurek:2013wia},  via interactions described by effective operators of the kind $\delta {\cal L}_{\rm asym} \sim {\cal O}_{\rm SM}^{+q} {\cal O}_{X}^{-q}$, where 
${\cal O}_{\rm SM}^{+q}$ contains fields that cascade down to SM particles and carries electric charge $q$, and ${\cal O}_{X}^{q}$  contains fields that cascade down to $X$ and carries electric charge $-q$, with $q\neq 0$. While $\delta {\cal L}_{\rm asym}$ conserves the total electric charge $Q \equiv Q_{\rm SM}  +  Q_{X}$, it violates the linear combination $A \equiv Q_{\rm SM}  -  Q_{X}$. It thus allows for the generation of a net number $\Delta A$, 
which implies the partition of the electric charge into two sectors, $\Delta Q_{\rm SM}  =  - \Delta Q_{X} = \Delta A/2$~\cite{Petraki:2013wwa}. A variety of mechanisms may be employed to create $\Delta A$, such as out-of-equilibrium decays of heavy particles~\cite{Falkowski:2011xh}, inelastic scatterings~\cite{Baldes:2014gca}, or the Affleck-Dine mechanism~\cite{Bell:2011tn}. Depending on the charge assignments of the fields that participate in $\delta {\cal L}_{\rm asym}$, a multi-level cascade of the electric charge down to the (millicharged) $X$ particles may be necessary, and can be realized  similarly as e.g. in Ref.~\cite{Gu:2010ft}.

\paragraph*{Case \#1} is where a millicharged $X$ makes a significant fraction of $\cO(50)$~MeV DM, and thus important cosmological and astrophysical constraints apply.   
If the $X$ species reaches chemical equilibrium in the early universe, then it must annihilate sufficiently, down to or below the observed DM density. The required annihilation cross-section is 
$\sigma_{\rm ann} v_{\rm rel} \gtrsim f_X^{-1}~6 \times 10^{-26}~{\rm cm^3/s}$.
The residual density of the $X$ antiparticles decreases nearly exponentially as the $\sigma_{\rm ann} v_{\rm rel}$ increases above the minimum required value, thus $\sigma_{\rm ann} v_{\rm rel}$ need not be very large~\cite{Graesser:2011wi}.
The $X$ species may annihilate into $e^+e^-$ pairs via an $s$-channel photon; however, the annihilation rate is inefficient. Sufficient annihilation may be ensured if the $X$ particles couple to and annihilate into lighter dark photons. The cosmological abundance of the dark photons may be subsequently depleted via decay into $e^+e^-$ pairs, if the dark photon mixes kinetically with the hypercharge gauge boson. The $X$ annihilation into dark photons has to deplete the $X$ antiparticle density below about $10^{-4}$ of the $X$ density, in order for the late-time residual annihilations to satisfy observational constraints~\cite{Baldes:2017gzw,Baldes:2017gzu}. The coupling of $X$ to the dark photon may also result in sizable DM self-scattering inside halos today that could explain current galactic structure observations~\cite{Tulin:2017ara}.
In the considered mass range, the relic $X$ population is evidently non-relativistic during BBN and recombination.

\paragraph*{Case \#2}  is similar to case  \#1 if $m_X > 1$~MeV, except that $X$ makes up only a small fraction of DM, $f_X \sim 10^{-3}$,  thus any cosmological and astrophysical constraints are relaxed. In the window $100$~keV-$1$~MeV, a freeze-in scenario similar to that described below for case \#4 may be operative 
(but with the constraints on the dark sector temperature eliminated, due to the larger $m_X$).

\paragraph*{Case \#3} 
{requires a particle with an electric charge $5 \lesssim \eps_X \lesssim 100$ and mass above 
$\sim 100$~GeV. 
For such a large charge, the early universe cosmology poses some challenge.  
Besides the stringent constraints from light element formation \cite{Pospelov:2006sc,Kohri:2006cn},  for $\eps_X \gg 1$ practically all $X$ particles would simply bind with protons, restoring $n_p = n_e$ and thus invalidating our mechanism.    
To prevent that, we introduce an unstable but long-lived particle $X'$ decaying to $X$ well after the BBN and before recombination, similarly as e.g. in Refs.~\cite{Bell:2010qt,Cline:2010kv}. 
For $Q_X' \gg 1$, $X'$ binds with protons before the BBN, which can  be broken by the late decays if the energy imparted to $X$ exceeds the binding energy.  
Note that  $X'$ annihilation into SM particles via its photon coupling  is Sommerfeld-enhanced, and is sufficient to deplete the symmetric component before the $X'$-$p$ pairs form.
One can also envisage more exotic realizations, e.g.~the $Q$-balls of Ref.~\cite{Shoemaker:2008gs} shedding the baryon charge on a long time scale. 
}

\paragraph*{Case \#4} deals with sub-eV particles with a tiny electric charge, $\eps_X \lesssim 10^{-14}$,  that may or may not constitute all of dark matter. 
If $X$ interacts very feebly and does not chemically equilibrate in the early universe, then the absence or sub-dominance of $X$ antiparticles may be inherited from its production. For example, the $X$ particles may be produced in the decays of a heavier species that carries the charge asymmetry at early times. The parent particles, $\hat{X}$, may possess a variety of interactions that deplete their antiparticles efficiently. Provided that  $\hat{X}$ decays into $X$ after the freeze-out of its annihilation processes, no significant density of $X$ antiparticles will be present today.  This implies $m_{\hat{X}}/ T_{\hat{X}}^{\rm prod} \gtrsim 20$, where we will allow for the temperature of $\hat{X}$ at the time of decay to differ from that of the SM plasma, by $\hat{\xi} \equiv T_{\hat{X}} / T_{\rm SM}$. 
In addition, the momentum imparted to the $X$ particles at their production has to be redshifted sufficiently at late times, to be consistent with conditions (iii) and (iv) above.  
The $X$ particles are produced with momentum $p_X^{\rm prod} \sim m_{\hat X}/2$ (or lower, for a many-body decay mode), which redshifts to $p_X = p_X^{\rm prod} \, (g_* / g_*^{\rm prod})^{1/3} \, (T_{\rm SM} / T_{\rm SM}^{\rm prod})$ at later times. 
Given the condition on the $T_{\hat{X}}^{\rm prod}$ mentioned above, we obtain
$p_X \gtrsim 3\hat{\xi} \, T_{\rm SM}$, where we set indicatively $(g_*/g_*^{\rm prod})^{1/3} \sim 1/3$. 
If the parent particles $\hat{X}$ were at a supercooled state with respect to the SM plasma at the time of their decay, such that  $\hat{\xi} \lesssim 10^{-2}$, then $X$ particles with $m_X \sim 0.1$~eV may be non-relativistic by the time of matter-radiation equality ($T_{\rm eq} \approx 0.8$~eV) and make up the DM of the universe (along the blue line in Fig.~\ref{fig:millicharge}). 
The same condition ensures that, if $f_X < 1$  (left of the blue line in Fig.~\ref{fig:millicharge}), the $X$ particles do not violate the constraints on extra radiation. 
Indeed, they contribute to the relativistic energy density of the universe by  $\rho_X \sim n_X p_X$, where $n_X = (1+z)^3 r_X \Omega_{b} \rho_c /m_p$. Adopting the $1\sigma$ constraint from Ref.~\cite{Ade:2015xua}, $\delta N_{\rm eff} \lesssim 0.334$, we shall require that 
$\rho_X \lesssim  0.05 \, T_{\rm SM}^4$.
This implies $\hat{\xi} \lesssim 10^{-2}$ for $\eps_X r_X \sim 3 \times 10^{-4}$ and $\epsilon_X \sim 10^{-14}$.

Finally, one may also envisage more exotic production mechanism leading to collective behavior of $X$,  e.g.~akin to the ``fuzzy" DM models of Refs.~\cite{Domcke:2014kla,Hui:2016ltb}.

{\bf Summary.} 
We have presented a novel mechanism to lower the baryon gas temperature during the cosmological dark ages.   
Our idea relies on the presence of a new, stable, negatively charged particle contributing  to the charge budget of the universe, such that $n_e \neq n_p$.  
The deficit of electrons during recombination results in an earlier decoupling of the baryon gas from the CMB, and thus a larger $T_\gamma/T_g$ ratio at the cosmic dawn.  
The parameter space  where the temperature ratio can be significantly enhanced is severely constrained by various measurements (distortion of the CMB spectrum, cooling of stars and supernovae, collider searches for exotic charged particles, etc.), cf.~\fref{millicharge}. 
However, a few allowed regions remain, notably sub-eV millicharged particles with $\eps_X \lesssim 10^{-14}$, MeV-scale millicharge particles that may or may not make all of dark matter,  or multi-charged $\cO(1)$~TeV particles.   
If this scenario is indeed responsible for the unexpectedly strong 21~cm absorption signal reported by the EDGES experiment, it can be tested in multiple ways. 
The baryon gas temperature is  significantly affected already at $z\sim 200$, cf. \fref{reco}, thus we predict a stronger absorption signal from the dark ages~\cite{Pritchard:2011xb}. 
Searches for exotic charged particles in collider experiments (at low energies as well as at the LHC) have a potential to close most of the remaining windows in the parameter space.    
The $X$ particles can also be searched in direct detection (especially in the millicharge case \cite{Essig:2017kqs}), and in cosmic rays (especially in the large charge regime \cite{Hu:2016xas}).  
Moreover, for $\epsilon_X \gtrsim 1$, $p$-$X$ pairs may be radiatively captured into atomic bound states  inside halos today, thus producing detectable signals in the X-rays~\cite{Belotsky:2014haa,Pearce:2015zca}.
Construction of complete, realistic models realizing our scenario and discussion of their observational consequences is postponed  to future publications.  

Aside from the anomaly in the EDGES result, the mere observation of the 21cm absorption signal can be interpreted as a constraint on new physics phenomena, e.g. on the annihilation cross section of dark matter \cite{DAmico:2018sxd}. In a similar fashion, our analysis can be re-interpreted as a bound on the cosmological abundance of new {\em positively} charged stable particles, as they would lead to a {\em suppression} of the absorption signal.

%%%%%%%%%%%%%%%%%%%%%%%%%%%%%%%%%%%%%%
\section*{Acknowledgements}
 
We thank Iason Baldes, Andreas Goudelis, Julia Harz, Alex Kusenko, Ian Low, Maxim Pospelov, and Michele Papucci  for useful discussions. 
A.F. is partially supported by the European Union's Horizon 2020
research and innovation programme under the Marie Sklodowska-Curie
grant agreements No 690575 and  No 674896.
K.P. was supported by the ANR ACHN 2015 grant (``TheIntricateDark" project), and by the NWO Vidi grant ``Self-interacting asymmetric dark matter".

%%%%%%%%%%
%\appendix
%\renewcommand{\theequation}{\Alph{section}.\arabic{equation}}  

%\section{Appendix title}
%\label{app:}

\bibliographystyle{JHEP}

\bibliography{as21_v2}

\providecommand{\href}[2]{#2}\begingroup\raggedright\begin{thebibliography}{10}

\bibitem{Bowman:2018yin}
J.~D. Bowman, A.~E.~E. Rogers, R.~A. Monsalve, T.~J. Mozdzen, and N.~Mahesh
  {\em Nature} {\bf 555} (2018), no.~7694 67--70.

\bibitem{Madau:1996cs}
P.~Madau, A.~Meiksin, and M.~J. Rees {\em Astrophys. J.} {\bf 475} (1997) 429,
  [\href{http://arxiv.org/abs/astro-ph/9608010}{{\tt astro-ph/9608010}}].

\bibitem{Zaldarriaga:2003du}
M.~Zaldarriaga, S.~R. Furlanetto, and L.~Hernquist {\em Astrophys. J.} {\bf
  608} (2004) 622--635, [\href{http://arxiv.org/abs/astro-ph/0311514}{{\tt
  astro-ph/0311514}}].

\bibitem{Ade:2015xua}
{\bf Planck} Collaboration, P.~A.~R. Ade et~al. {\em Astron. Astrophys.} {\bf
  594} (2016) A13, [\href{http://arxiv.org/abs/1502.01589}{{\tt
  arXiv:1502.01589}}].

\bibitem{AliHaimoud:2010dx}
Y.~Ali-Haimoud and C.~M. Hirata {\em Phys. Rev.} {\bf D83} (2011) 043513,
  [\href{http://arxiv.org/abs/1011.3758}{{\tt arXiv:1011.3758}}].

\bibitem{Fraser:2018acy}
S.~Fraser et~al. \href{http://arxiv.org/abs/1803.03245}{{\tt
  arXiv:1803.03245}}.

\bibitem{Pospelov:2018kdh}
M.~Pospelov, J.~Pradler, J.~T. Ruderman, and A.~Urbano
  \href{http://arxiv.org/abs/1803.07048}{{\tt arXiv:1803.07048}}.

\bibitem{Dvorkin:2013cea}
C.~Dvorkin, K.~Blum, and M.~Kamionkowski {\em Phys. Rev.} {\bf D89} (2014),
  no.~2 023519, [\href{http://arxiv.org/abs/1311.2937}{{\tt arXiv:1311.2937}}].

\bibitem{Tashiro:2014tsa}
H.~Tashiro, K.~Kadota, and J.~Silk {\em Phys. Rev.} {\bf D90} (2014), no.~8
  083522, [\href{http://arxiv.org/abs/1408.2571}{{\tt arXiv:1408.2571}}].

\bibitem{Barkana:2018lgd}
R.~Barkana {\em Nature} {\bf 555} (2018), no.~7694 71--74,
  [\href{http://arxiv.org/abs/1803.06698}{{\tt arXiv:1803.06698}}].

\bibitem{Munoz:2018pzp}
J.~B. Munoz and A.~Loeb \href{http://arxiv.org/abs/1802.10094}{{\tt
  arXiv:1802.10094}}.

\bibitem{Berlin:2018sjs}
A.~Berlin, D.~Hooper, G.~Krnjaic, and S.~D. McDermott
  \href{http://arxiv.org/abs/1803.02804}{{\tt arXiv:1803.02804}}.

\bibitem{Barkana:2018qrx}
R.~Barkana, N.~J. Outmezguine, D.~Redigolo, and T.~Volansky
  \href{http://arxiv.org/abs/1803.03091}{{\tt arXiv:1803.03091}}.

\bibitem{Peebles:1968ja}
P.~J.~E. Peebles {\em Astrophys. J.} {\bf 153} (1968) 1.

\bibitem{Zeldovich:1969en}
{\relax Ya}.~B. Zeldovich, V.~G. Kurt, and R.~A. Sunyaev {\em Sov. Phys. JETP}
  {\bf 28} (1969) 146. [Zh. Eksp. Teor. Fiz.55,278(1968)].

\bibitem{Senatore:2008vi}
L.~Senatore, S.~Tassev, and M.~Zaldarriaga {\em JCAP} {\bf 0908} (2009) 031,
  [\href{http://arxiv.org/abs/0812.3652}{{\tt arXiv:0812.3652}}].

\bibitem{AliHaimoud:2010ab}
Y.~Ali-Haimoud and C.~M. Hirata {\em Phys. Rev.} {\bf D82} (2010) 063521,
  [\href{http://arxiv.org/abs/1006.1355}{{\tt arXiv:1006.1355}}].

\bibitem{Dolgov:2013una}
A.~D. Dolgov, S.~L. Dubovsky, G.~I. Rubtsov, and I.~I. Tkachev {\em Phys. Rev.}
  {\bf D88} (2013), no.~11 117701, [\href{http://arxiv.org/abs/1310.2376}{{\tt
  arXiv:1310.2376}}].

\bibitem{Vinyoles:2015khy}
N.~Vinyoles and H.~Vogel {\em JCAP} {\bf 1603} (2016), no.~03 002,
  [\href{http://arxiv.org/abs/1511.01122}{{\tt arXiv:1511.01122}}].

\bibitem{Chang:2018rso}
J.~H. Chang, R.~Essig, and S.~D. McDermott
  \href{http://arxiv.org/abs/1803.00993}{{\tt arXiv:1803.00993}}.

\bibitem{Prinz:1998ua}
A.~A. Prinz et~al. {\em Phys. Rev. Lett.} {\bf 81} (1998) 1175--1178,
  [\href{http://arxiv.org/abs/hep-ex/9804008}{{\tt hep-ex/9804008}}].

\bibitem{Soper:2014ska}
D.~E. Soper, M.~Spannowsky, C.~J. Wallace, and T.~M.~P. Tait {\em Phys. Rev.}
  {\bf D90} (2014), no.~11 115005, [\href{http://arxiv.org/abs/1407.2623}{{\tt
  arXiv:1407.2623}}].

\bibitem{Davidson:2000hf}
S.~Davidson, S.~Hannestad, and G.~Raffelt {\em JHEP} {\bf 05} (2000) 003,
  [\href{http://arxiv.org/abs/hep-ph/0001179}{{\tt hep-ph/0001179}}].

\bibitem{Chatrchyan:2013oca}
{\bf CMS} Collaboration, S.~Chatrchyan et~al. {\em JHEP} {\bf 07} (2013) 122,
  [\href{http://arxiv.org/abs/1305.0491}{{\tt arXiv:1305.0491}}].

\bibitem{Aad:2015oga}
{\bf ATLAS} Collaboration, G.~Aad et~al. {\em Eur. Phys. J.} {\bf C75} (2015)
  362, [\href{http://arxiv.org/abs/1504.04188}{{\tt arXiv:1504.04188}}].

\bibitem{Petraki:2013wwa}
K.~Petraki and R.~R. Volkas {\em Int. J. Mod. Phys.} {\bf A28} (2013) 1330028,
  [\href{http://arxiv.org/abs/1305.4939}{{\tt arXiv:1305.4939}}].

\bibitem{Zurek:2013wia}
K.~M. Zurek {\em Phys. Rept.} {\bf 537} (2014) 91--121,
  [\href{http://arxiv.org/abs/1308.0338}{{\tt arXiv:1308.0338}}].

\bibitem{Falkowski:2011xh}
A.~Falkowski, J.~T. Ruderman, and T.~Volansky {\em JHEP} {\bf 05} (2011) 106,
  [\href{http://arxiv.org/abs/1101.4936}{{\tt arXiv:1101.4936}}].

\bibitem{Baldes:2014gca}
I.~Baldes, N.~F. Bell, K.~Petraki, and R.~R. Volkas {\em Phys. Rev. Lett.} {\bf
  113} (2014), no.~18 181601, [\href{http://arxiv.org/abs/1407.4566}{{\tt
  arXiv:1407.4566}}].

\bibitem{Bell:2011tn}
N.~F. Bell, K.~Petraki, I.~M. Shoemaker, and R.~R. Volkas {\em Phys. Rev.} {\bf
  D84} (2011) 123505, [\href{http://arxiv.org/abs/1105.3730}{{\tt
  arXiv:1105.3730}}].

\bibitem{Gu:2010ft}
P.-H. Gu, M.~Lindner, U.~Sarkar, and X.~Zhang {\em Phys.Rev.} {\bf D83} (2011)
  055008, [\href{http://arxiv.org/abs/1009.2690}{{\tt arXiv:1009.2690}}].

\bibitem{Graesser:2011wi}
M.~L. Graesser, I.~M. Shoemaker, and L.~Vecchi {\em JHEP} {\bf 10} (2011) 110,
  [\href{http://arxiv.org/abs/1103.2771}{{\tt arXiv:1103.2771}}].

\bibitem{Baldes:2017gzw}
I.~Baldes and K.~Petraki {\em JCAP} {\bf 1709} (2017), no.~09 028,
  [\href{http://arxiv.org/abs/1703.00478}{{\tt arXiv:1703.00478}}].

\bibitem{Baldes:2017gzu}
I.~Baldes, M.~Cirelli, P.~Panci, K.~Petraki, F.~Sala, and M.~Taoso
  \href{http://arxiv.org/abs/1712.07489}{{\tt arXiv:1712.07489}}.

\bibitem{Tulin:2017ara}
S.~Tulin and H.-B. Yu {\em Phys. Rept.} {\bf 730} (2018) 1--57,
  [\href{http://arxiv.org/abs/1705.02358}{{\tt arXiv:1705.02358}}].

\bibitem{Pospelov:2006sc}
M.~Pospelov {\em Phys. Rev. Lett.} {\bf 98} (2007) 231301,
  [\href{http://arxiv.org/abs/hep-ph/0605215}{{\tt hep-ph/0605215}}].

\bibitem{Kohri:2006cn}
K.~Kohri and F.~Takayama {\em Phys. Rev.} {\bf D76} (2007) 063507,
  [\href{http://arxiv.org/abs/hep-ph/0605243}{{\tt hep-ph/0605243}}].

\bibitem{Bell:2010qt}
N.~F. Bell, A.~J. Galea, and R.~R. Volkas {\em Phys. Rev.} {\bf D83} (2011)
  063504, [\href{http://arxiv.org/abs/1012.0067}{{\tt arXiv:1012.0067}}].

\bibitem{Cline:2010kv}
J.~M. Cline, A.~R. Frey, and F.~Chen {\em Phys. Rev.} {\bf D83} (2011) 083511,
  [\href{http://arxiv.org/abs/1008.1784}{{\tt arXiv:1008.1784}}].

\bibitem{Shoemaker:2008gs}
I.~M. Shoemaker and A.~Kusenko {\em Phys. Rev.} {\bf D78} (2008) 075014,
  [\href{http://arxiv.org/abs/0809.1666}{{\tt arXiv:0809.1666}}].

\bibitem{Domcke:2014kla}
V.~Domcke and A.~Urbano {\em JCAP} {\bf 1501} (2015), no.~01 002,
  [\href{http://arxiv.org/abs/1409.3167}{{\tt arXiv:1409.3167}}].

\bibitem{Hui:2016ltb}
L.~Hui, J.~P. Ostriker, S.~Tremaine, and E.~Witten {\em Phys. Rev.} {\bf D95}
  (2017), no.~4 043541, [\href{http://arxiv.org/abs/1610.08297}{{\tt
  arXiv:1610.08297}}].

\bibitem{Pritchard:2011xb}
J.~R. Pritchard and A.~Loeb {\em Rept. Prog. Phys.} {\bf 75} (2012) 086901,
  [\href{http://arxiv.org/abs/1109.6012}{{\tt arXiv:1109.6012}}].

\bibitem{Essig:2017kqs}
R.~Essig, T.~Volansky, and T.-T. Yu {\em Phys. Rev.} {\bf D96} (2017), no.~4
  043017, [\href{http://arxiv.org/abs/1703.00910}{{\tt arXiv:1703.00910}}].

\bibitem{Hu:2016xas}
P.-K. Hu, A.~Kusenko, and V.~Takhistov {\em Phys. Lett.} {\bf B768} (2017)
  18--22, [\href{http://arxiv.org/abs/1611.04599}{{\tt arXiv:1611.04599}}].

\bibitem{Belotsky:2014haa}
K.~Belotsky, M.~Khlopov, C.~Kouvaris, and M.~Laletin {\em Adv. High Energy
  Phys.} {\bf 2014} (2014) 214258, [\href{http://arxiv.org/abs/1403.1212}{{\tt
  arXiv:1403.1212}}].

\bibitem{Pearce:2015zca}
L.~Pearce, K.~Petraki, and A.~Kusenko {\em Phys. Rev.} {\bf D91} (2015) 083532,
  [\href{http://arxiv.org/abs/1502.01755}{{\tt arXiv:1502.01755}}].

\bibitem{DAmico:2018sxd}
G.~D'Amico, P.~Panci, and A.~Strumia
  \href{http://arxiv.org/abs/1803.03629}{{\tt arXiv:1803.03629}}.

\end{thebibliography}\endgroup

\end{document}